\documentclass[conference]{IEEEtran}
\usepackage{cite}
\ifCLASSINFOpdf
  \usepackage[pdftex]{graphicx}
\else
\fi
\usepackage[cmex10]{amsmath}
\usepackage{url}
\usepackage[flushleft]{threeparttable}

\begin{document}
\title{Assessing Progress of Parkinson's Disease Using Acoustic Analysis of Phonation}


\author{\IEEEauthorblockN{Jiri Mekyska\IEEEauthorrefmark{1},
Zoltan Galaz\IEEEauthorrefmark{1},
Zdenek Mzourek\IEEEauthorrefmark{1},
Zdenek Smekal\IEEEauthorrefmark{1},
Irena Rektorova\IEEEauthorrefmark{2}\IEEEauthorrefmark{3},
Ilona Eliasova\IEEEauthorrefmark{2}\IEEEauthorrefmark{3},\\
Milena Kostalova\IEEEauthorrefmark{3}\IEEEauthorrefmark{4},
Martina Mrackova\IEEEauthorrefmark{2}\IEEEauthorrefmark{3},
Dagmar Berankova\IEEEauthorrefmark{3},
Marcos Faundez-Zanuy\IEEEauthorrefmark{5},\\
Karmele L\'{o}pez-de-Ipi\~{n}a\IEEEauthorrefmark{6} and
Jesus B. Alonso-Hernandez\IEEEauthorrefmark{7}}
\IEEEauthorblockA{\IEEEauthorrefmark{1}Department of Telecommunications\\
Brno University of Technology, Technicka~10, 61600~Brno, Czech Republic\\
Email: mekyska@feec.vutbr.cz}
\IEEEauthorblockA{\IEEEauthorrefmark{2}First Department of Neurology\\
St. Anne's University Hospital, Pekarska~53, 65691~Brno, Czech Republic}
\IEEEauthorblockA{\IEEEauthorrefmark{3}Applied Neuroscience Research Group\\
Central European Institute of Technology, Masaryk University, Komenskeho~nam.~2, 60200~Brno, Czech Republic}
\IEEEauthorblockA{\IEEEauthorrefmark{4}Department of Neurology\\
Faculty Hospital and Masaryk University, Jihlavska 20, 63900~Brno, Czech Republic}
\IEEEauthorblockA{\IEEEauthorrefmark{5}Escola Universitaria Politecnica de Mataro\\
Tecnocampus, Avda. Ernest Lluch~32, 08302~Mataro, Barcelona, Spain}
\IEEEauthorblockA{\IEEEauthorrefmark{6}Department of Systems Engineering and Automation\\
University of the Basque Country UPV/EHU, Av de Tolosa 54, 20018~Donostia, Spain}
\IEEEauthorblockA{\IEEEauthorrefmark{7}Institute for Technological Development and Innovation in Communications (IDeTIC)\\
University of Las Palmas de Gran Canaria, 35001 Las Palmas de Gran Canaria, Spain}}


\maketitle

\begin{abstract}
%\boldmath
This paper deals with a~complex acoustic analysis of phonation in patients with Parkinson's disease (PD) with a~special focus on estimation of disease progress that is described by 7 different clinical scales (e.\,g. Unified Parkinson's disease rating scale or Beck depression inventory). The analysis is based on parametrization of 5 Czech vowels pronounced by 84 PD patients. Using classification and regression trees we estimated all clinical scores with maximal error lower or equal to 13\,\%. Best estimation was observed in the case of Mini-mental state examination (MAE = 0.77, estimation error 5.50\,\%). Finally, we proposed a~binary classification based on random forests that is able to identify Parkinson's disease with sensitivity SEN = 92.86\,\% (SPE = 85.71\,\%). The parametrization process was based on extraction of 107 speech features quantifying different clinical signs of hypokinetic dysarthria present in PD.
\end{abstract}


% \fi
\IEEEpeerreviewmaketitle

\section{Introduction}
Parkinson's disease (PD) is the second most frequent neurodegenerative disease caused by a progressive loss of dopaminergic neurons (particularly in the substancia nigra pars compacta), however it can also result from anti-psychotic medications or frequent blows to head~\cite{Skodda2010}. In 60\,--\,90\,\% of PD patients aged over 65 years it is possible to observe speech disorder called hypokinetic dysarthria (HD)~\cite{Sapir2008}. HD manifests itself in areas of phonation, articulation, prosody, fluency and faciokinesis. Speech and voice disturbances are characterized by monotonous pitch and loudness, decreased stress and emphasis, breathy and harsh voice, reduced vocal intensity, variable rates including short rushes of speech or accelerated speech, consonant imprecision, impaired breath support for speech, reduction in phonation time, difficulty in the initiation of speech activities, and inappropriate pausing~\cite{Goberman2005b, Skodda2009, Mekyska2011b, Eliasova2013}.

Clinical neurologists and psychologists use different scales to rate PD. Some of them were developed for complex assessment of PD (e.\,g. UPDRS\,--\,Unified Parkinson's disease rating scale~\cite{Fahn1987}, NMSS\,--\,Non-motor symptoms scale~\cite{Chaudhuri2007}) and some of them are focused on specific clinical signs like depression (e.\,g. BDI\,--\,Beck depression inventory~\cite{Beck1961}), dyskinesia (e.\,g. AIMS\,--\,Abnormal involuntary movements scale~\cite{Rush2000}), gait (e.\,g. FOG\,--\,Freezing of gait questionnaire~\cite{Giladi2000}), sleep disorders (e.\,g. RBDSQ\,--\,The REM sleep behavior disorder screening questionnaire~\cite{Stiasny2007}) or cognitive impairments (e.\,g. MMSE\,--\,Mini-mental state examination~\cite{Folstein1975}, ACE-R\,--\,Addenbrooke's cognitive examination-revised~\cite{Larner2007}). Although significance of these scores have been robustly validated on large sets of patients, they have one disadvantage and that is the factor of subjectivity during examination. In all the cases we should count with possibility that two independent clinicians can rate the same patient with different scores.

Therefore scientists started to deal with new paraclinical methods of PD assessment that are objective, non-invasive, quick, low-cost and that significantly correlate with conventional rating scales. It is not considered that these methods will substitute clinicians during diagnosis and rating, they should rather bring new kind of biomarkers and generally parameters that will be used for objective and quick estimation of progress and that can be used for precise disease monitoring in time.

One of the paraclinical methods is the acoustic analysis of dysarthric speech. Just a limited number of works focused on assessment Parkinson's disease exists in this field of science. Moreover scientists usually deal only with UPDRS (mainly part III: motor examination; part V: Modified Hoehn and Yahr Staging)~\cite{Tsanas2010b, Tsanas2012b, Asgari2010, Skodda2012b, Tsanas2010, Howard2013, Eskidere2012}. Score estimation of the other, and still important, rating scales has not been investigated yet. Therefore, in this work, we aim to: 1) identify vowels whose analysis provides best estimation of particular clinical scores used for assessment of PD; 2) introduce a new concept of Parkinson's disease progress quantification based on acoustic analysis of phonation; 3) propose a binary classifier of PD.

The rest of this paper is organized as follows. Sections \ref{sec:data} and \ref{sec:methodology} describe the dataset and methodology respectively. Section \ref{sec:results} provides some preliminary results where we firstly performed binary classification (parkinsonic people vs. healthy controls) and Spearman's correlation between clinical scores and particular speech features. Next, we employed classification and regression trees in order to achieve best estimation in terms of low mean absolute correlation and high Pearson's correlation coefficient. The conclusion is given in sec.\,\ref{sec:conclusion}.

\section{Data}
\label{sec:data}

We included in this study 84 PD patients (36 women, 48 men) and 49 (24 women, 25 men) age and gender matched healthy controls (HC) who were enrolled at the First Department of Neurology, St. Anne's University Hospital in Brno, Czech Republic. The healthy participants had no history or presence of brain diseases (including neurological and psychiatric illnesses) or speech disorders. For more demographic characteristics of the PD group, see Table~\ref{tab:demographic}. All participants signed an informed consent form that had been approved by the Ethics Committee of St. Anne's University Hospital in Brno.

\begin{table}
		\caption{Demographic and clinical characteristics of PD patients}
		\label{tab:demographic}
		\centering
		\begin{threeparttable}
		\begin{tabular}{l c c}
		\hline
		\hline
		Speakers & PD (females) & PD (males)\\
		\hline
		Number & 36 & 48\\
		Age (years) & 68.47 $\pm$ 7.64 & 66.21 $\pm$ 8.78\\
		PD duration (years) & 7.61 $\pm$ 4.85 & 7.83 $\pm$ 4.39\\
		UPDRS III & 22.06 $\pm$ 13.73 & 26.85 $\pm$ 10.22\\
		UPDRS IV & 2.72 $\pm$ 3.01 & 3.15 $\pm$ 2.59\\
		RBDSQ & 3.42 $\pm$ 3.48 & 3.85 $\pm$ 2.99\\
		FOG & 6.94 $\pm$ 5.72 & 6.67 $\pm$ 5.57\\
		NMSS & 36.03 $\pm$ 26.72 & 38.19 $\pm$ 19.72\\
		BDI & 18.57 $\pm$ 23.94 & 9.69 $\pm$ 6.23\\
		MMSE & 27.38 $\pm$ 3.63 & 28.56 $\pm$ 1.05\\
		ACE-R & 87.00 $\pm$ 8.62 & 88.08 $\pm$ 7.16\\
		LED (mg) & 862.44 $\pm$ 508.3 & 1087.00 $\pm$ 557.47\\
		\hline
		\hline
		\end{tabular}
		\begin{tablenotes}
      \item[1] UPDRS III\,--\,Unified Parkinson's disease rating scale, part III: Motor Examination; UPDRS IV\,--\,Unified Parkinson's disease rating scale, part IV: Complications of Therapy; RBDSQ\,--\,The REM sleep behavior disorder screening questionnaire); FOG\,--\,Freezing of gait questionnaire; NMSS\,--\,Non-motor symptoms scale; BDI\,--\,Beck depression inventory; MMSE\,--\,Mini-mental state examination; ACE-R\,--\,Addenbrooke's cognitive cxamination-revised; LED\,--\,L-dopa equivalent daily dose
    \end{tablenotes}
		\end{threeparttable}
\end{table}

Each of the participants was firstly examined by clinical neurologist and psychologist who rated him according to 8 scales: Unified Parkinson's disease rating scale (part III and IV), The REM sleep behavior disorder screening questionnaire, Freezing of gait questionnaire, Non-motor symptoms scale, Beck depression inventory, Mini-mental state examination and Addenbrooke's cognitive cxamination-revised. In addition we processed duration of disease and LED\,--\,L-dopa equivalent daily dose.

After the clinical examination the participants uttered 4 sets of five Czech vowels ([a], [e], [i], [o], [u]): 1) s\,--\,short vowels pronounced with normal intensity; 2) l\,--\,sustained vowels pronounced with normal intensity; 3) ll\,--\,sustained vowels pronounced with maximum intensity; 4) ls\,--\,sustained vowels pronounced with minimum intensity, but not whispered. 

\section{Methodology}
\label{sec:methodology}

Speech samples were digitized by sampling frequency $f_{\mathrm{s}} = 48\,\mbox{kHz}$ and consequently resampled to 16\,kHz. The recordings were parametrized by NDAT (Neurological Disorder Analysis Tool)~\cite{Eliasova2013, Mekyska2011c}, developed at the Brno University of Technology, and statistically processed in MATLAB.

\subsection{Parametrization}

The feature extraction process included calculation of wide range of speech features that we divided into several groups (description of all below mentioned features can be found in our recent article~\cite{Mekyska2015}):
\begin{enumerate}
	\item Features describing phonation\,--\,$F_{0}$, 5 kinds of jitter and 6 kinds of shimmer~\cite{Praat}, PPE (Pitch Period Entropy)~\cite{Little2009}, a~measure of standard deviation (std) of the time that vocal folds are apart ($\mbox{GQ}_{\mathrm{open}}$) and in collisions respectively ($\mbox{GQ}_{\mathrm{closed}}$)~\cite{Tsanas2010}, $E$ (short-time energy), TKEO (Teager-Kaiser Energy Operator)~\cite{Dimitriadis2009}, ME (4\,Hz modulation energy)~\cite{Falk2012}, MPSD (Median of Power Spectral Density)~\cite{Gonzalez2010} and LSTER (Low Short-Time Energy Ratio)~\cite{Song2009}.
	\item Features describing tongue movement\,--\,formants $F_{1}$--$F_{3}$ and their bandwidths $BW_{1}$--$BW_{3}$, VSA (vowel space area)~\cite{Sapir2010} and lnVSA (its logarithmic version)~\cite{Sapir2010}, FCR (formant centralization ratio)~\cite{Sapir2010}, VAI (vowel articulation index)~\cite{Skodda2010} and $F_{2\mathrm{i}}/F_{2\mathrm{u}}$ (ratio of second formants of vowels [i] and [u]).
	\item Features describing speech quality\,--\,ZCR (Zero-Crossing Rate), HZCRR (High Zero-Crossing Rate Ratio)~\cite{Song2009}, FLUF (Fraction of Locally Unvoiced Frames)~\cite{Alonso2001}, SF (Spectral Flux)~\cite{Banchhor2012}, SDBM (Spectral Distance Based on Module), SDBP (Spectral Distance Based on Phase)~\cite{Alonso2001}, CPP (Cepstral Peak Prominence)~\cite{Hillenbrand1996}, PECM (Pitch Energy Cepstral Measure)~\cite{Alonso2001}, VR (Variation in Ratio between the second/first harmonic within the derived cepstral domain)~\cite{Alonso2001}, HNR (Harmonic-to-Noise Ratio), NHR (Noise-to-Harmonic Ratio), NNE (Normalized Noise Energy)~\cite{Kasuya1986}, GNE (Glottal-to-Noise Excitation ratio)~\cite{Michaelis1997}, SPI (Soft Phonation Index)~\cite{Deliyski1993}, VTI (Voice Turbolence Index)~\cite{Deliyski1993}, SSD (Segmental Signal-to-Dysperiodicity ratio), MSER (Modulation Spectra Energy Ratio), MFP (Modulation Frequency of Peak)~\cite{Alpan2011},  RPHM (Relative Peak Height of Modulation spectra)~\cite{Mekyska2015}, ICER (Inferior Colliculus Energy Ratio)~\cite{Mekyska2015} and RPHIC (Relative Peak Height of Inferior Colliculus)~\cite{Mekyska2015}.
	\item Features based on bispectrum~\cite{Alonso2001} and bicepstrum~\cite{Mekyska2015}\,--\,BII (Bicoherence Index Interference), HFEB (High Frequency Energy of one-dimensional Bicoherence), LFEB (low Frequency Energy of one-dimensional Bicoherence), BMII (Bispectrum Module Interference Index), BPII (Bispectrum Phase Interference Index), BCII (BiCepstral Index Interference), HFEBC (High Frequency Energy of one-dimensional BiCepstral index), LFEBC (Low Frequency Energy of one-dimensional BiCepstral index), CMII (BiCepstrum Module Interference Index), BCPII (BiCepstrum Phase Interference Index), LCBCER (Low Cepstra/BiCepstra Energy Ratio), HCBCER (High Cepstra/BiCepstra Energy Ratio), LSBER (Low Spectra/Bispectra Energy Ratio), HSBER (High Spectra/Bispectra Energy Ratio), BCMD (BiCepstral Module Distance) and BCPD (BiCepstral Phase Distance).
	\item Features based on empirical mode decomposition~\cite{Mekyska2015,Tsanas2010}\,--\, $\mbox{IMF-SNR}_\mathrm{TKEO}$ (SNR based on Teager-Kaiser Energy Operator extracted from intrinsic mode functions), $\mbox{IMF-SNR}_\mathrm{SEO}$ (based on Squared Energy Operator), $\mbox{IMF-SNR}_\mathrm{SE}$ (based on Shannon Entropy), $\mbox{IMF-SNR}_\mathrm{RE}$ (based on second-order R{\'e}nyi Entropy), $\mbox{IMF-SNR}_\mathrm{ZCR}$ (based on Zero-Crossing Rate), $\mbox{IMF-NSR}_\mathrm{TKEO}$, $\mbox{IMF-NSR}_\mathrm{SEO}$, $\mbox{IMF-NSR}_\mathrm{SE}$, $\mbox{IMF-NSR}_\mathrm{RE}$, IMF-FD (Fractal Dimension extracted from the $1^{\mathrm{st}}$ IMF), IMF-CPP (Cepstral Peak Prominence extracted from the $1^{\mathrm{st}}$ IMF) and IMF-GNE (Glottal-to-Noise Excitation ratio based on the $1^{\mathrm{st}}$ IMF).
	\item Non-linear dynamic features\,--\,CD (Correlation Dimension)~\cite{Vaziri2010}, FD (Fractal Dimension)~\cite{Vaziri2010}, ZL (Ziv-Lempel complexity)~\cite{Aboy2006}, HE (Hurst Exponent)~\cite{Orozco2011}, SHE (Shannon Entropy), RE (second-order R{\'e}nyi Entropy), CE (Correlation Entropy)~\cite{Jayawardena2010}, RBE1 (first-order R{\'e}nyi Block Entropy)~\cite{Henriquez2009}, RBE2 (second-order R{\'e}nyi Block Entropy)~\cite{Henriquez2009}, AE (Approximate Entropy) and SE (Sample Entropy)~\cite{Yentes2013} with 8 different kernels~\cite{Mekyska2015}, PE (Permutation Entropy), FMMI (First Minimum of Mutual Information function)~\cite{Henriquez2009} and LLE (Largest Lyapunov Exponent)~\cite{Orozco2011}.
\end{enumerate}

If the feature has been represented by vector, we have employed transformation to scalar value based on median, standard deviation (std), $1^{\mathrm{st}}$ percentile (1p), 99th percentile (99p) and interpercentile range (ir) defined as 99p -- 1p. Considering all the possible combinations, we have extracted approximately 350 features for each vowel.

\subsection{Preliminary analysis}

To accomplish the $3^{\mathrm{rd}}$ goal of this work (binary classification) we firstly employed RF (random forests) classifier along with sequential forward feature selection (SFFS). Precision of PD identification was tested in terms of classification accuracy (ACC), sensitivity (SEN), specificity (SPE) and trade-off between sensitivity and specificity (TSS) defined as:
\begin{eqnarray}
\mbox{TSS} = 2^{\sin\left(\frac{\pi\cdot \mathrm{SEN}}{2}\right)\sin\left(\frac{\pi\cdot \mathrm{SPE}}{2}\right)}.
\end{eqnarray}
We considered two scenarios: 1) individual vowel analysis; 2) classification within each vowel set (see sec.\,\ref{sec:data}). In both cases we used leave-one-out validation.

Next, we performed Spearman's rank correlation between particular feature vector and clinical information (PD duration, UPDRS III and IV, RBDSQ, FOG, NMSS, BDI, MMSE\, ACE-R and LED) in order to find a possible candidate for preliminary PD assessment.

\begin{table}[h]
		\caption{Classification results}
		\label{tab:resall}
		\centering
		\begin{threeparttable}
		\begin{tabular}{l c c c c c}
		\hline
		\hline
		Vowels & ACC [\%] & SEN [\%] & SPE [\%] & TSS & No.\\
		\hline
		a (s) & 75.19 & 79.76 & 67.35 & 1.7748 & 2\\
		e (s) & 79.70 & 78.57 & 81.63 & 1.8724 & 7\\
		i (s) & 83.46 & 84.52 & 81.63 & 1.9059 & 9\\  
		o (s) & 87.22 & 89.29 & 83.67 & 1.9367 & 11\\
		u (s) & 79.70 & 79.76 & 79.59 & 1.8680 & 5\\
		\hline
		a (l) & 73.68 & 71.43 & 77.55 & 1.7969 & 6\\
		e (l) & \textbf{88.72} & \textbf{91.67} & \textbf{83.67} & \textbf{1.9440} & 8\\
		i (l) & 81.95 & 83.33 & 79.59 & 1.8878 & 8\\
		o (l) & 72.93 & 72.62 & 73.47 & 1.7791 & 2\\
		u (l) & 76.69 & 78.57 & 73.47 & 1.8189 & 7\\
		\hline
		a (ll) & 75.94 & 77.38 & 73.47 & 1.8116 & 7\\
		e (ll) & 83.46 & 86.90 & 77.55 & 1.8904 & 10\\
		i (ll) & 74.44 & 73.81 & 75.51 & 1.8020 & 3\\
		o (ll) & 81.95 & 83.33 & 79.59 & 1.8878 & 9\\
		u (ll) & 75.19 & 78.57 & 69.39 & 1.7861 & 3\\
		\hline
		a (ls) & 73.68 & 77.38 & 67.35 & 1.7616 & 2\\
		e (ls) & 78.20 & 77.38 & 79.59 & 1.8529 & 5\\
		i (ls) & 81.20 & 83.33 & 77.55 & 1.8745 & 5\\
		o (ls) & 84.96 & 83.33 & 87.76 & 1.9293 & 3\\
		u (ls) & 81.95 & 86.90 & 73.47 & 1.8598 & 7\\
		\hline
		all (s) & 81.20 & 83.33 & 77.55 & 1.8745 & 3\\
		all (l) & 84.21 & 82.14 & 87.76 & 1.9228 & 9\\
		all (ll) & 83.46 & 83.33 & 83.67 & 1.9110 & 3\\
		all (ls) & \textbf{90.23} & \textbf{92.86} & \textbf{85.71} & \textbf{1.9572} & 11\\
		\hline
		\hline
		\end{tabular}
		\begin{tablenotes}
      \item[1] ACC\,--\,classification accuracy; SEN\,--\,sensitivity; SPE\,--\,specificity; TSS\,--\,trade-off between sensitivity and specificity; No.\,--\,number of selected features; s\,--\,short vowel pronounced with normal intensity; l\,--\,sustained vowel pronounced with normal intensity; ll\,--\,sustained vowel pronounced with maximum intensity; ls\,--\,sustained vowel pronounced with minimum intensity (not whispering)
    \end{tablenotes}
		\end{threeparttable}
\end{table}

\begin{figure*}[t!]
	\centering
		\includegraphics[width=1.00\textwidth]{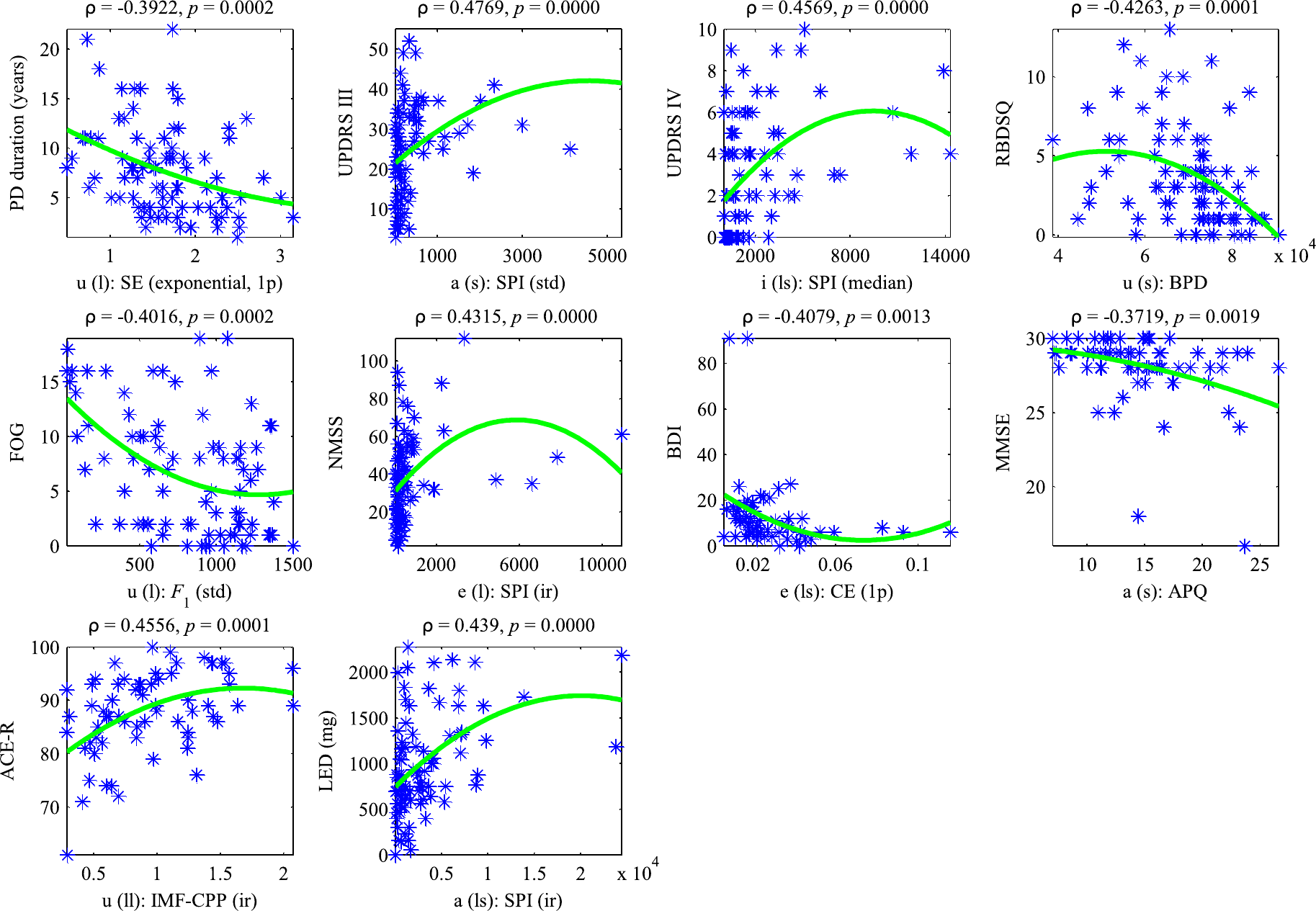}
	\caption{Most significant correlations between clinical and paraclinical data (UPDRS III\,--\,Unified Parkinson's disease rating scale, part III: Motor Examination; UPDRS IV\,--\,Unified Parkinson's disease rating scale, part IV: Complications of Therapy; RBDSQ\,--\,The REM sleep behavior disorder screening questionnaire); FOG\,--\,Freezing of gait questionnaire; NMSS\,--\,Non-motor symptoms scale; BDI\,--\,Beck depression inventory; MMSE\,--\,Mini-mental state examination; ACE-R\,--\,Addenbrooke's cognitive examination-revised; LED\,--\,L-dopa equivalent daily dose; $\rho$\,--\,Spearman's correlation coefficient; $p$\,--\,significance level of correlation; s\,--\,short vowel pronounced with normal intensity; l\,--\,sustained vowel pronounced with normal intensity; ll\,--\,sustained vowel pronounced with maximum intensity; ls\,--\,sustained vowel pronounced with minimum intensity (not whispering))}
	\label{fig:corrgraphs}
\end{figure*}

\begin{table}
		\caption{Lowest estimation errors of clinical scores}
		\label{tab:selres}
		\centering
		\begin{threeparttable}
		\begin{tabular}{l c c c c c c}
		\hline
		\hline
		Clin. info. & Range & Vowel & MAE & EE1 [\%] & EE2 [\%]\\
		\hline
		PD duration & 0\,--\,not lim. & e (s) & 2.25 & 10.71 & --\\
		UPDRS III & 0\,--\,108 & i (ll) & 5.70 & 10.96 & 5.28\\
		UPDRS IV & 0\,--\,23 & all (ls) & 1.30 & 13.00 & 5.65\\
		RBDSQ & 0\,--\,13 & u (l) & 1.54 & 11.85 & 11.85\\
		FOG & 0\,--\,24 & all (s) & 2.30 & 12.11 & 9.58\\
		NMSS & 0\,--\,360 & all (ls) & 11.48 & 10.44 & 3.19\\
		BDI & 0\,--\,63 & u (l) & 3.12 & 8.43 & 4.95\\
		MMSE & 0\,--\,30 & o (l) & 0.77 & 5.50 & 2.57\\
		ACE-R & 0\,--\,100 & all (ll) & 3.58 & 9.18 & 3.58\\
		LED (mg) & 0\,--\,not lim. & i (l) & 224.99 & 9.89 & --\\
		\hline
		\hline
		\end{tabular}
		\begin{tablenotes}
      \item[1] UPDRS III\,--\,Unified Parkinson's disease rating scale, part III: Motor Examination; UPDRS IV\,--\,Unified Parkinson's disease rating scale, part IV: Complications of Therapy; RBDSQ\,--\,The REM sleep behavior disorder screening questionnaire); FOG\,--\,Freezing of gait questionnaire; NMSS\,--\,Non-motor symptoms scale; BDI\,--\,Beck depression inventory; MMSE\,--\,Mini-mental state examination; ACE-R\,--\,Addenbrooke's cognitive examination-revised; LED\,--\,L-dopa equivalent daily dose; MAE\,--\,mean absolute error; EE1\,--\,estimation error, type 1; EE2\,--\,estimation error, type 2
    \end{tablenotes}
		\end{threeparttable}
\end{table}

\subsection{Classification and regression}

After the preliminary analysis we moved further and trained classification and regression trees (CART) to estimate selected clinical information with higher accuracy measured by mean absolute error (MAE) and Pearson's correlation coefficient $\rho$. We used a~two-step feature selection. Firstly we preselected 500 features using filtering method based on minimum redundancy and maximum relevance (mRMR). Then we employed wrapping method based on SFFS to select the final feature subset. We considered the same two scenarios as in the previous analysis: 1) individual vowel analysis; 2) classification within each vowel set. We used the leave-one-out validation in both cases.

Finally we selected for each clinical information the best combination of vowels and speech features and calculated two kinds of estimation error to better describe ability of PD progress assessment. We measured these two estimation errors:
\begin{eqnarray}
\mbox{EE1}&=&\frac{\mathrm{MAE}}{\mathrm{range}(\mathrm{CI})},\\
\mbox{EE2}&=&\frac{\mathrm{MAE}}{\max(\mathrm{CI})},
\end{eqnarray}
where CI stands for particular clinical information (e.\,g. FOG). Function $\mathrm{range}(\mathrm{CI})$ calculates the range from clinical data available during the analysis, while function $\max(\mathrm{CI})$ returns the maximal score that can be theoretically reached in the specific scale.

\section{Experimental results}
\label{sec:results}

\begin{table*}
		\caption{Estimation of clinical scores}
		\label{tab:resregress}
		\centering
		\begin{threeparttable}
		\begin{tabular}{l	c	c	c	c	c	c	c	c	c	c	c	c	c	c c}
		\hline
		\hline
		& \multicolumn{3}{c}{PD duration (years)} & \multicolumn{3}{c}{UPDRS III} & \multicolumn{3}{c}{UPDRS IV} & \multicolumn{3}{c}{RBDSQ} & \multicolumn{3}{c}{FOG}\\
		Vowels & MAE & $\rho$ & No. & MAE & $\rho$ & No. & MAE & $\rho$ & No. & MAE & $\rho$ & No. & MAE & $\rho$ & No.\\
		\hline
		a (s) & 3.21 & 0.4253 & 4 & 8.28 & 0.4956 & 7 & 1.42 & 0.6690 & 7 & 1.88 & 0.6002 & 7 & 3.32 & 0.5903 & 7 \\
		e (s) & \textbf{2.25} & \textbf{0.6891} & 19 & 6.61 & 0.7084 & 6 & 1.47 & 0.6600 & 15 & 2.09 & 0.6117 & 2 & 3.86 & 0.4692 & 7 \\
		i (s) & 2.89 & 0.5511 & 8 & 8.33 & 0.3934 & 14 & 1.59 & 0.5695 & 7 & 2.09 & 0.4493 & 3 & 3.99 & 0.4748 & 6 \\
		o (s) & 3.28 & 0.4860 & 2 & 7.59 & 0.5444 & 8 & 1.90 & 0.3706 & 3 & 1.83 & 0.6436 & 10 & 3.55 & 0.5479 & 8 \\
		u (s) & 3.13 & 0.4059 & 7 & 8.18 & 0.5121 & 4 & 1.69 & 0.5835 & 6 & 2.14 & 0.4644 & 6 & 4.42 & 0.3792 & 2 \\
		a (l) & 3.31 & 0.3952 & 1 & 8.02 & 0.5153 & 5 & 1.60 & 0.5926 & 7 & 2.23 & 0.4138 & 5 & 4.03 & 0.4104 & 5 \\
		e (l) & 2.40 & 0.6856 & 9 & 6.34 & 0.7210 & 11 & 1.87 & 0.5092 & 6 & 1.83 & 0.6034 & 8 & 3.40 & 0.6190 & 5 \\
		i (l) & 2.93 & 0.6077 & 3 & 8.62 & 0.4570 & 5 & 2.10 & 0.4181 & 4 & 1.83 & 0.5679 & 13 & 3.43 & 0.6131 & 3 \\
		o (l) & 2.38 & 0.6802 & 10 & 7.20 & 0.5874 & 5 & 2.01 & 0.3882 & 2 & 2.23 & 0.3710 & 2 & 3.64 & 0.4920 & 5 \\
		u (l) & 3.00 & 0.5580 & 3 & 8.44 & 0.5367 & 3 & 1.62 & 0.5958 & 6 & \textbf{1.54} & \textbf{0.7358} & 8 & 3.90 & 0.4861 & 5 \\
		a (ll) & 2.75 & 0.6185 & 7 & 6.59 & 0.6540 & 7 & 2.00 & 0.4339 & 5 & 2.47 & 0.2644 & 1 & 4.22 & 0.3571 & 3 \\
		e (ll) & 2.95 & 0.5214 & 7 & 7.32 & 0.6340 & 7 & 1.76 & 0.5336 & 5 & 1.79 & 0.5931 & 9 & 4.29 & 0.3551 & 1 \\
		i (ll) & 3.23 & 0.4638 & 5 & \textbf{5.70} & \textbf{0.7987} & 15 & 1.94 & 0.4579 & 4 & 1.86 & 0.6190 & 6 & 2.67 & 0.7409 & 19 \\
		o (ll) & 2.60 & 0.6527 & 9 & 6.72 & 0.6696 & 15 & 1.85 & 0.4738 & 4 & 1.92 & 0.5799 & 9 & 3.21 & 0.6097 & 11 \\
		u (ll) & 3.18 & 0.4933 & 7 & 8.03 & 0.5401 & 5 & 1.75 & 0.5824 & 7 & 1.85 & 0.6242 & 5 & 3.83 & 0.5074 & 2 \\
		a (ls) & 2.54 & 0.6806 & 11 & 7.52 & 0.6078 & 6 & 1.45 & 0.7059 & 10 & 2.35 & 0.3313 & 2 & 3.28 & 0.5803 & 4 \\
		e (ls) & 3.53 & 0.1794 & 1 & 9.41 & 0.3471 & 2 & 1.90 & 0.5051 & 4 & 2.10 & 0.5335 & 3 & 3.73 & 0.5206 & 6 \\
		i (ls) & 2.78 & 0.5168 & 9 & 8.84 & 0.4878 & 2 & 1.42 & 0.6689 & 12 & 2.19 & 0.4407 & 2 & 3.04 & 0.6527 & 7 \\
		o (ls) & 2.69 & 0.5940 & 11 & 7.50 & 0.6224 & 9 & 1.48 & 0.6105 & 6 & 1.70 & 0.5441 & 14 & 4.06 & 0.4493 & 1 \\
		u (ls) & 3.56 & 0.3257 & 2 & 7.27 & 0.6827 & 13 & 1.64 & 0.5695 & 7 & 1.70 & 0.6473 & 11 & 4.40 & 0.3483 & 3 \\
		all (s) & 2.49 & 0.6290 & 12 & 6.74 & 0.6339 & 12 & 1.60 & 0.5816 & 7 & 2.06 & 0.4971 & 2 & \textbf{2.30} & \textbf{0.8031} & 11 \\
		all (l) & 2.68 & 0.5312 & 7 & 7.22 & 0.6169 & 5 & 1.60 & 0.6164 & 6 & 2.25 & 0.3697 & 1 & 2.49 & 0.7723 & 10 \\
		all (ll) & 3.29 & 0.2752 & 3 & 7.50 & 0.6163 & 6 & 1.38 & \textbf{0.6850} & 12 & 1.82 & 0.5719 & 12 & 3.11 & 0.6416 & 10 \\
		all (ls) & 2.60 & 0.5751 & 12 & 8.07 & 0.5449 & 5 & \textbf{1.30} & 0.6768 & 12 & 1.72 & 0.6912 & 7 & 3.75 & 0.4187 & 7 \\
		\hline
		& & & & & & & & & & & & & & &\\
				& \multicolumn{3}{c}{NMSS} & \multicolumn{3}{c}{BDI} & \multicolumn{3}{c}{MMSE} & \multicolumn{3}{c}{ACE-R} & \multicolumn{3}{c}{LED (mg)}\\
		Vowels & MAE & $\rho$ & No. & MAE & $\rho$ & No. & MAE & $\rho$ & No. & MAE & $\rho$ & No. & MAE & $\rho$ & No.\\
		\hline
		a (s) & 16.16 & 0.3712 & 4 & 5.25 & 0.8991 & 5 & 1.31 & 0.3171 & 3 & 5.50 & 0.4435 & 1 & 373.17 & 0.4993 & 4 \\
		e (s) & 15.09 & 0.5613 & 6 & 4.13 & 0.9361 & 7 & 1.02 & 0.7297 & 5 & 4.29 & 0.6942 & 7 & 331.00 & 0.5706 & 11 \\
		i (s) & 13.75 & 0.5552 & 7 & 5.22 & 0.8962 & 5 & 0.92 & 0.8475 & 10 & 4.04 & 0.7266 & 5 & 334.89 & 0.5950 & 6 \\
		o (s) & 14.14 & 0.6116 & 8 & 5.47 & 0.7689 & 4 & 1.48 & 0.1309 & 1 & 5.60 & 0.4200 & 1 & 308.26 & 0.6622 & 14 \\
		u (s) & 13.61 & 0.6322 & 7 & 5.86 & 0.7195 & 9 & 1.33 & 0.6302 & 4 & 4.86 & 0.5733 & 4 & 373.20 & 0.5471 & 5 \\
		a (l) & 14.42 & 0.6185 & 4 & 5.46 & 0.8856 & 7 & 0.83 & 0.7975 & 9 & 5.35 & 0.2980 & 3 & 389.50 & 0.3761 & 3 \\
		e (l) & 14.27 & 0.5747 & 4 & 4.39 & 0.9319 & 5 & 1.25 & 0.2195 & 11 & 4.98 & 0.5204 & 2 & 302.08 & 0.6705 & 16 \\
		i (l) & 12.81 & 0.6482 & 9 & 4.58 & 0.9301 & 8 & 1.21 & 0.4788 & 1 & 4.35 & 0.6023 & 9 & \textbf{224.99} & \textbf{0.8232} & 11 \\
		o (l) & 12.23 & 0.6985 & 8 & 5.95 & 0.6026 & 7 & \textbf{0.77} & 0.8601 & 8 & 4.01 & 0.6609 & 8 & 376.21 & 0.4955 & 3 \\
		u (l) & 17.07 & 0.3669 & 3 & \textbf{3.12} & \textbf{0.9634} & 7 & 1.08 & 0.7038 & 4 & 4.62 & 0.5431 & 7 & 365.32 & 0.4624 & 6 \\
		a (ll) & 14.57 & 0.5123 & 3 & 4.59 & 0.7066 & 8 & 1.28 & 0.4743 & 4 & 4.10 & 0.6519 & 9 & 269.60 & 0.7004 & 7 \\
		e (ll) & 17.60 & 0.3268 & 2 & 5.67 & 0.7163 & 8 & 0.94 & 0.7346 & 11 & 4.04 & 0.6814 & 12 & 410.26 & 0.3478 & 4 \\
		i (ll) & 12.88 & 0.5513 & 7 & 7.04 & 0.3014 & 8 & 1.11 & 0.7182 & 6 & 3.68 & 0.7107 & 9 & 344.12 & 0.5468 & 6 \\
		o (ll) & 16.19 & 0.4495 & 6 & 6.92 & 0.2276 & 7 & 1.17 & 0.5647 & 5 & 5.10 & 0.4611 & 4 & 414.54 & 0.3817 & 4 \\
		u (ll) & 13.74 & 0.5636 & 12 & 3.86 & 0.9414 & 9 & 1.17 & 0.5554 & 9 & 4.06 & 0.7242 & 6 & 396.35 & 0.4483 & 4 \\
		a (ls) & 16.37 & 0.4265 & 2 & 8.03 & 0.6593 & 2 & 1.33 & 0.4625 & 2 & 4.28 & 0.6101 & 6 & 318.72 & 0.7018 & 6 \\
		e (ls) & 15.05 & 0.4814 & 3 & 4.19 & 0.9396 & 5 & 1.11 & 0.6857 & 10 & 4.98 & 0.5322 & 7 & 391.48 & 0.4532 & 6 \\
		i (ls) & 15.70 & 0.4662 & 3 & 6.87 & 0.4022 & 3 & 0.93 & 0.8075 & 10 & 5.30 & 0.3864 & 3 & 360.22 & 0.4983 & 6 \\
		o (ls) & 14.68 & 0.5298 & 10 & 6.23 & 0.7116 & 2 & 1.46 & 0.0085 & 1 & 4.10 & 0.6554 & 12 & 289.96 & 0.6944 & 10 \\
		u (ls) & 15.43 & 0.4327 & 4 & 3.76 & 0.9434 & 9 & 1.46 & 0.1606 & 1 & 5.35 & 0.5067 & 5 & 337.65 & 0.5836 & 9 \\
		all (s) & 12.43 & 0.5946 & 10 & 5.73 & 0.7469 & 5 & 0.81 & 0.8400 & 12 & 4.04 & 0.6424 & 15 & 281.46 & 0.7002 & 11 \\
		all (l) & 13.16 & 0.6029 & 12 & 3.61 & 0.9502 & 7 & 0.84 & \textbf{0.8725} & 6 & 3.87 & 0.7318 & 11 & 268.04 & 0.7313 & 11 \\
		all (ll) & 15.00 & 0.4796 & 10 & 3.84 & 0.9391 & 8 & 0.94 & 0.7192 & 9 & \textbf{3.58} & \textbf{0.7694} & 10 & 315.85 & 0.6560 & 7 \\
		all (ls) & \textbf{11.48} & \textbf{0.7190} & 16 & 6.34 & 0.5648 & 9 & 1.38 & 0.6490 & 2 & 4.55 & 0.5685 & 6 & 367.90 & 0.5306 & 4 \\
		\hline
		\hline
		\end{tabular}
		\begin{tablenotes}
      \item[1] UPDRS III\,--\,Unified Parkinson's disease rating scale, part III: Motor Examination; UPDRS IV\,--\,Unified Parkinson's disease rating scale, part IV: Complications of Therapy; RBDSQ\,--\,The REM sleep behavior disorder screening questionnaire); FOG\,--\,Freezing of gait questionnaire; NMSS\,--\,Non-motor symptoms scale; BDI\,--\,Beck depression inventory; MMSE\,--\,Mini-mental state examination; ACE-R\,--\,Addenbrooke's cognitive examination-revised; LED\,--\,L-dopa equivalent daily dose; MAE\,--\,mean absolute error; $\rho$\,--\,Pearson's correlation coefficient; No.\,--\,number of selected features; s\,--\,short vowel pronounced with normal intensity; l\,--\,sustained vowel pronounced with normal intensity; ll\,--\,sustained vowel pronounced with maximum intensity; ls\,--\,sustained vowel pronounced with minimum intensity (not whispering)
    \end{tablenotes}
		\end{threeparttable}
\end{table*}

Results from the binary classification can be found in Table~\ref{tab:resall}. In the first scenario (individual vowel analysis) the best results were observed in the case of classification based on parametrization of sustained vowel [e] (ACC = 88.72\,\%, SEN = 91.67\,\%, SPE = 83.67\,\%, TSS = 1.9440\,\%). 8 features were selected by SFFS. Although most of the researchers use sustained vowel [a] in order to diagnose PD from speech, no clear explanation for this selection has been published yet. One possible reason is that tongue goes to maximal vertical position during pronunciation of this vowel (see Hellwag triangle~\cite{Mol1965}), but on the other hand there is nearly no movement in horizontal one. Other publications also show that significance of vowel [a] in PD patients should not be dogma~\cite{Bolanos2013, Orozco2013, Orozco2013b}. Therefore it is still necessary to come up with a~robust and complex work that will clarify the significance of particular vowels based on testing on multilingual PD databases (theoretically the effect of culture and language on phonation is minimal, but this must be proved in the case of dysarthric speech as well).

In the second scenario we considered classification within each vowel set. The results show that the best discrimination power provide features extracted from sustained vowels pronounced with minimum intensity (ACC = 90.23\,\%, SEN = 92.86\,\%, SPE = 85.71\,\%, TSS = 1.9572\,\%, number of selected features: 11). This result hits another dogma and that is the analysis of sustained vowels pronounced with normal intensity. In our recent paper we have proved that sustained vowels pronounced with minimum intensity (not whispering) accent vocal tremor and they are more complicated for precise vocal fold vibration (more than in the case of normal sustained vowels where a speaker does not have to concentrate too much on precise voicing\,--\,he couldn't whisper)~\cite{Smekal2015}.

Next, we tried to find a possible candidate for preliminary PD assessment using Spearman's rank correlation between particular feature vector and selected clinical information. We have visualized the results using correlation graphs that can be seen on Fig.\,\ref{fig:corrgraphs}. Each graph contains a~non-linear regression line defined by second-order polynomial. Although all correlation are significant ($p < 0.01$), the correlation coefficients in absolute values are not so high and it is clear that it is necessary to include more features in order to reach better clinical score estimation. 

Therefore, in the next step, we employed CART along with SFFS. The obtained results are given in Table~\ref{tab:resregress}. Finally, to better evaluate accuracy of estimation we selected for each clinical information the best MAE and additionally computed EE1 and EE2 (see Table~\ref{tab:selres}). Regarding EE1 the lowest estimation error was observed in the case of Mini-mental state examination (possible range: 0\,--\,30, MAE = 0.77 and resulting EE1 = 5.50\,\%) and the highest one in the case of UPDRS IV (possible range: 0\,--\,23, MAE = 1.30, EE1 = 13.00\,\%). This type of error is probably more relevant, because it is quantified using the range of data that were really processed. However, we also provide estimation accuracy in terms of EE2 that is related to the maximal value that can be in specific scale reached. From this point of view the lowest estimation error was observed in the case of MMSE too (EE2 = 2.57\,\%), but the highest one was measured in the case of RBDSQ (possible range: 0\,--\,13, MAE = 1.54, EE1 = 11.85\,\%).

\section{Conclusion}
\label{sec:conclusion}

In this paper we performed a~complex acoustic analysis of phonation in patients with Parkinson's disease in order to estimate a~degree of this disease measured by 7 different clinical scales. The analysis is base on parametrization of 5 Czech vowels pronounced by 84 PD patients. We achieved all goals of this work: 1) We identified vowels that are suitable for estimation of particular clinical scores (see Table~\ref{tab:selres}). 2) We proposed a concept of PD progress assessment that is based on CART and features coming from different domains (description of phonation, tongue movement, speech quality, bispectrum, bicepstrum, empirical mode decomposition, etc.). We evaluated the proposed concept using estimation of 10 different clinical scores (PD duration, UPDRS III and IV, RBDSQ, FOG, NMSS, BDI, MMSE\, ACE-R and LED). All scores were estimated with maximal estimation error lower or equal to 13\,\%. Best estimation was observed in the case of Mini-mental state examination (EE1 = 5.50\,\%). 3) We proposed a~binary classification based on random forests that is able to identify Parkinson's disease with sensitivity SEN = 92.86\,\% (SPE = 85.71\,\%). The classifier is fed by features extracted from sustained vowels pronounced with minimum intensity.

All the considered scores are given by clinicians who examine the patients subjectively. Therefore it would be interesting to calculate difference of estimation among approximately 10 clinical neurologists/psychologists and compare it to estimation error that was measured in this work. Other interesting idea for future work would be to deeper investigate relations between speech and depression or sleep disorders. It is well documented that speech disorders in PD patients significantly correlate with changes in gait~\cite{Moreau2007}. However, according to estimation errors of BDI or RBDSQ (which are lower than in the case of FOG), we can suppose that some hidden relations between depression, sleeps disorders and speech are far more stronger.

\section*{Acknowledgment}
Research described in this paper was financed by the National Sustainability Program under grant LO1401 and by projects NT13499 (Speech, its impairment and cognitive performance in Parkinson's disease), COST IC1206, project ``CEITEC, Central European Institute of Technology'': (CZ.1.05/1.1.00/02.0068) from the European Regional Development Fund, FEDER and Ministerio de Econom\'{i}a y Competitividad TEC2012-38630-C04-03 (Kingdom of Spain). For the research, infrastructure of the SIX Center was used.



%\bibliographystyle{IEEEtran}
%\begin{thebibliography}{1}
  %0.5em minus 0.4em\relax Harlow, England: Addison-Wesley, 1999.
%
%\end{thebibliography}

\bibliographystyle{IEEEtran}
\bibliography{IWOBI2015}

\end{document}